\documentclass[twocolumn,floatfix,eqsecnum,rmp]{revtex4}
\usepackage{graphicx}
\usepackage{amsmath}
\usepackage{bm}
\bibpunct{}{}{,}{s}{}{\textsuperscript{,}}
\newcommand{\be}{\begin{equation}}
\newcommand{\ee}{\end{equation}}

\begin{document}
\title{Quantum Point Contacts and Coherent Electron Focusing}
\author{H. van Houten and C. W. J. Beenakker}
\affiliation{Philips Research Laboratories, 5600 JA Eindhoven, The Netherlands}
\begin{abstract}
The theory of quantum ballistic transport, applied to quantum point contacts and coherent electron focusing in a two-dimensional electron gas, is reviewed in relation to experimental observations, stressing its character of electron optics in the solid state. It is proposed that an optical analogue of the conductance quantization of quantum point contacts can be constructed, and a theoretical analysis is presented. Coherent electron focusing is discussed as an experimental realization of mode-interference in ballistic transport.\bigskip\\
{\tt Published in {\em Analogies in Optics and Micro Electronics},\\
edited by W. van Haeringen and D. Lenstra (Kluwer, Dordrecht, 1990).}
\end{abstract}
\maketitle

\tableofcontents

\section{\label{sec1} Introduction}

Electron optics as an experimental discipline\cite{ref1} can be traced back to Busch\cite{ref2} who demonstrated the focusing action of an axial magnetic field on electron beams in vacuum. The similarity of the properties of ballistic conduction electrons in a degenerate electron gas and those of free electrons in vacuum suggests the possibility of electron optics in the solid state. Classical ballistic transport in metals, which has the character of geometrical optics, has been realized with the pioneering work of Sharvin\cite{ref3} and Tsoi\cite{ref4} on point contacts and electron focusing. This work has recently been extended to the quantum ballistic transport regime in the two-dimensional electron gas\footnote{A 2DEG is a degenerate electron gas which is strongly confined in one directon by an electrostatic potential well at the interface between two semiconductors, such that only the lowest quantum level in the well is occupied. Electrons in a 2DEG are thus dynamically constrained to move in a plane, and accordingly there is a 2D density of states.\cite{ref5}}
(2DEG) in a GaAs--AlGaAs heterostructure.\cite{ref6,ref7,ref8} 
Essential advantages of this system are its reduced dimensionality and lower electron gas density, with correspondingly large Fermi wavelength, which can be varied locally by means of gate electrodes. Point contacts of variable width of the order of the Fermi wavelength have been defined by applying a negative voltage to a split-gate on top of the heterostructure. 

\begin{figure}
\centerline{\includegraphics[width=8cm]{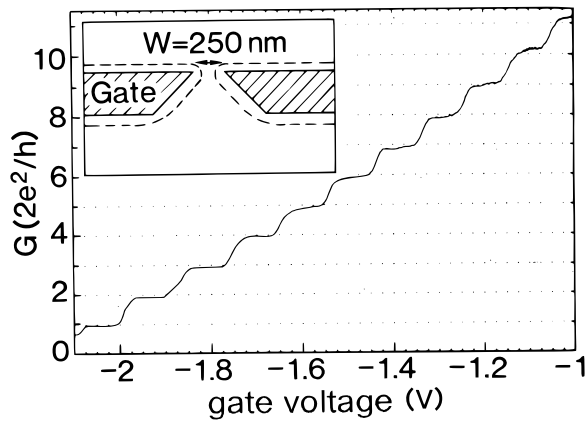}}
\caption{
Conductance of a quantum point contact in the absence of a magnetic field. The inset shows the sample geometry schematically. The point contact is electrostatically defined as a constriction in the two-dimensional electron gas. The constriction width incrases continuously as the (negative) gate voltage is decreased, but the conductance increases in steps given by $2e^{2}/h$. [From Ref.\ \cite{ref6}]
\label{fig1}
}
\end{figure}

An interesting and unexpected finding was the quantization of the conductance of these {\em quantum point contacts\/} in units of $2e^{2}/h$ (see Fig. \ref{fig1}).\cite{ref6,ref7} This new effect can be understood by the similarity of transport through the point contact and propagation through an electron waveguide. Point contacts can also be used to inject a divergent beam of ballistic electrons in the 2DEG. The focusing of such a beam by a transverse magnetic field, acting as a {\em lens}, has been demonstrated in an electron focusing experiment.\cite{ref8,ref9}
The boundary of the 2DEG, also defined by means of a gate, acts as a high quality {\em mirror}, causing specular boundary scattering.\cite{ref8,ref9} For sufficiently large gate voltages the point contact width is smaller than the Fermi wavelength of the electrons. Quantum point contacts in this regime act as monochromatic {\em point sources},\cite{ref10} as demonstrated by the large interference structure found experimentally.\cite{ref8,ref9} The first building blocks for the exploitation of solid state electron optics have thus been realized.

In this chapter we discuss the theory of quantum ballistic transport applied to the conductance through quantum point contacts\cite{ref6,ref7} and coherent electron focusing,\cite{ref8,ref9,ref10} stressing its character of electron optics in the solid state. A review with a wider scope and more experimental detail is Ref.\ \cite{ref11}. Reviews of related topics in quantum ballistic transport can be found in Refs.\ \cite{ref12,ref13,ref14}. Section \ref{sec4} of this paper, dealing with the optical analogue of the conductance quantization of a
quantum point contact, does not contain previously published material.

\section{\label{sec2} Electrons at the Fermi level}

In this section we discuss some elementary properties of the single electron states
in a degenerate two-dimensional electron gas. In a large system, and in the
effective mass approximation, these states are plane waves with wave vector $\mathbf{k}$.
The conduction band is approximated by
\be
E(\mathbf{k})=\frac{\hbar^{2}k^{2}}{2m},   \label{eq1}
\ee
with $m$ the effective mass, which for GaAs is $0.067\,m_{\rm e}$. Linear transport at low temperatures can be formulated in terms of motion of electrons at the Fermi level $E_{\rm F}$. The group velocity of these conduction electrons is $\mathbf{v}_{\mathrm{F}}=\partial E/\hbar\mathbf{k}=\hbar\mathbf{k}_{\mathrm{F}}/m$,
and their wavelength $\lambda_{\mathrm{F}}=h/m v_{\rm F}$ is $\approx 40\,\mathrm{n}\mathrm{m}$ in the experiments. These quantities
are obtained from the electron gas density $n_{\rm s}=(k_{\mathrm{F}})^{2}/2\pi$, which is a directly
measurable quantity.

In a channel with width comparable to $\lambda_{\mathrm{F}}$ the transverse motion of the
electrons is quantized, while the motion along the channel is free. The wave
function is separable in a transverse bound state with quantum number n, and a
longitudinal plane wave $\exp(iky)$. The dispersion relation $E_{\mathrm{n}}(k)$ of these quasi
one-dimensional subbands or transverse waveguide modes is
\be
E_{\mathrm{n}}(k)=E_{\mathrm{n}}(0)+\frac{\hbar^{2}k^{2}}{2m},  \label{eq2}
\ee
where $E_{\rm n}(0)$ is the energy of the n-th bound state. The frequency $E_{\rm n}(0)/\hbar$ is the cut-off frequency of the mode, as in an optical fiber. The number of occupied
modes $N$ is the largest integer n such that $E_{\mathrm{n}}(0)\leq E_{\mathrm{F}}$. Since $E$ is still quadratic in
$k$, the group velocity $\hbar k/m$ remains linear in the wave number, as in the bulk
2DEG.

This changes if we apply a magnetic field $B$ perpendicular to the electron gas.
Let the channel be in the $y$-direction, and $B$ in the $z$-direction, so that the 2DEG
is in the $x-y$ plane. In the Landau gauge $\mathbf{A} =(0,Bx,0)$ the wave function remains
separable as in the absence of a magnetic field. However, $E_{\mathrm{n}}(k)$ is no longer
quadratic in $k$, so that the group velocity differs from $\hbar k/m$. Note that it is the
group velocity which is relevant for conduction through the channel. In fact, the
current carried by a mode n is proportional to the product of the group velocity $v_{\mathrm{n}}=dE_{\mathrm{n}}(k)/\hbar dk$ and the density of states $\rho_{\mathrm{n}}$, both evaluated at the Fermi energy.
Since the number of states per unit channel length in an interval $dk$ is $2\times dk/2\pi$
(with an additional factor of 2 from the spin degeneracy) the energy density of
states is $\rho_{\mathrm{n}}=(\pi dE_{\mathrm{n}}(k)/dk)^{-1}$. It follows that $\rho_{\mathrm{n}}v_{\mathrm{n}}=2/h$ is {\it independent\/} of wave
number or mode index, regardless of the form of the dispersion relation.
Formulated differently, this tells us that in an electron waveguide the {\it current is shared equally among the modes}. This is the basic reason for the conductance
quantization discussed in the following section.

\section{\label{sec3} Conductance quantization of a quantum point contact}

Van Wees {\em et al.}\cite{ref6} and Wharam {\em et al.}\cite{ref7} observed a sequence of steps in the conductance $G$ of a point contact as its width was varied (by means of a gate voltage, see Fig.\ \ref{fig1}). The steps were at integer multiples of $2e^{2}/h\approx (13\,{\rm k}\Omega)^{-1}$, a combination of fundamental constants which is familiar from the quantum Hall effect.\cite{ref15} However, the conductance quantization was observed in the {\em absence\/} of a magnetic field, as well as in the presence of a magnetic field. An elementary explanation of this effect relies on the fact that the point contact acts in a way as an electron waveguide or multi-mode fiber.\cite{ref16,ref17} Each populated one-dimensional subband or transverse waveguide mode contributes $2e^{2}/h$ to the conductance because of the cancellation of the group velocity and the one-dimensional density of states discussed in sec.\ \ref{sec2}. Since the number $N$ of occupied modes is necessarily an integer, it follows from this simple argument that the total conductance is quantized
\be
G=\frac{2e^{2}}{h}N,\label{eq3}
\ee
as observed experimentally.\footnote{The resistance $1/G$ is non-zero for an ideal ballistic conductor of finite width, because it is a {\em contact\/} resistance. The existence of a contact resistance of order $h/Ne^{2}$ was first pointed out by Imry.\cite{ref18}}

For a square well confinement potential of width $W$ one has $N=k_{\rm F}W/\pi$ in zero magnetic field, if $W\gg\lambda_{\rm F}$ so that the discreteness of $N$ may be ignored. Eq.\ (\ref{eq3}) then gives the 2D analogue of the 3D Sharvin formula \cite{ref3} for the conductance of a classical ballistic point contact. However, the above argument leading to (\ref{eq3}) holds irrespective of the shape of the confinement potential, or of the presence of a magnetic field.

A more detailed explanation of the conductance quantization requires a consideration of the coupling between quantum states in the narrow point contact to those in the wide 2DEG regions. As discussed in sec.\ \ref{sec6} in the more general context of electron focusing, this is essentially a transmission problem. Here it suffices to say that if the point contact opens up gradually into the wide regions, the transport is {\em adiabatic}\cite{ref19,ref20,ref21} from the entrance to the exit of the point contact. The lowest $N$ subbands at the wide entrance region (in one-to-one correspondence to those occupied in the point contact) are transmitted with probability 1, the others being reflected.

For the case that the point contact widens {\em abruptly\/} into the 2DEG, deviations from (\ref{eq3}) occur, for short channels mainly because of evanescent waves (or modes with a frequency above the cut-off frequency) which have a non-zero transmission probability, for longer channels mainly because of quantum mechanical reflections at their entrance and exit. Extensive numerical and analytical work\cite{ref22,ref23,ref24,ref25,ref26} has demonstrated that (\ref{eq3}) is still a surprisingly good approximation --- although transmission resonances may obscure the plateaus. Experimentally, a limited increase in temperature helps to smooth out the resonances and to improve the flatness of the plateaus.

\section{\label{sec4} Optical analogue of the conductance quantization}

The analogy of ballistic transport through quantum point contacts with
transmission of light through a multi mode fiber, or microwaves through a
waveguide (summarized in Table \ref{table1}), naturally triggers the question: ``Does the
quantized conductance have an optical analogue, and if it does, why was it not
known?'' In this section we want to show that: ``Yes, there is an optical analogue,
but the experiment is not as natural for light as it is for electrons.''

\begin{table}
\caption{
\label{table1}}
\begin{ruledtabular}
\begin{tabular}{ll}
photons& electrons\\
\hline
ray& trajectory\\
mode& subband\\
mode index& quantum number $n$\\
wave number $k$& canonical momentum $\hbar k$\\
frequency $\omega$& energy $E=\hbar\omega$\\
dispersion law $\omega(k)$& bandstructure $E_{n}(k)$\\
group velocity $d\omega/dk$& group velocity $dE/\hbar dk$
\end{tabular}
\end{ruledtabular}
\end{table}

Consider monochromatic light, of frequency $\omega$ and polarization
$E_{\mathrm{x}}=E_{\mathrm{y}}=B_{\mathrm{z}}=0$, incident on a long slit (along the $z$-axis) in a metallic screen.
The non-zero electric field component $E_{\mathrm{z}}(x,y)$ then satisfies the two-dimensional
scalar wave equation (Ref.\ \cite{ref27}, page 561)
\be
\frac{\partial^{2}E_{\mathrm{z}}}{\partial x^{2}} + \frac{\partial^{2}E_{\mathrm{z}}}{\partial y^{2}}+k^{2}E_{\rm z}=0,\label{eq4}
\ee
with $k\equiv\omega/c\equiv 2\pi/\lambda$ the wave number ($c$ is the velocity of light, and $\lambda$ the
wavelength). This equation is identical to the Schr\"{o}dinger equation for the wave
function $\Psi(x,y)$ of an electron at the Fermi level in a 2DEG, with the identification $k\equiv k_{\rm F}$. If the boundaries of the 2DEG are modeled by infinite potential walls, then 
the boundary conditions are also the same in the two problems ($\Psi$ and $E_{\mathrm{z}}$ both
vanish for $(x,y)$ on the boundary). From $E_{\mathrm{z}}$ one finds the non-zero magnetic field
components $B_{\mathrm{x}}=-(\mathrm{i}/k)\partial E_{\mathrm{z}}/\partial y$, and $B_{\mathrm{y}}=-(\mathrm{i}/k)\partial E_{\mathrm{z}}/\partial_{\mathrm{x}}$, and hence the energy flux
\be
\mathbf{j}=\frac{c}{8\pi}\mathrm{R}\mathrm{e}(\mathbf{E}\times\mathbf{B}^*)=\frac{c}{8\pi k}\mathrm{R}\mathrm{e}(\mathrm{i}E_{\mathrm{z}}\nabla E_{\mathrm{z}}^*).   \label{eq5}
\ee
This expression is identical, up to a numerical factor, to the quantum mechanical
expression for the particle flux, $(\hbar/m)\mathrm{R}\mathrm{e}(\mathrm{i}\Psi\nabla\Psi^{*})$. It follows that the ratio of
transmitted to incident power in the optical problem is the same as the ratio of the
transmitted to incident current in its electronic counterpart.

In optics one usually studies the transmission of a single incident plane wave,
as a function of the angle of incidence. In a 2DEG, however, electrons are incident
from all directions in the $x-y$ plane, with an isotropic velocity distribution. The
incident flux then has a $\cos\theta$ angular distribution, where $\theta$ is the angle with the
normal to the screen. [Such an isotropic distribution results naturally from a
``good'' electron reservoir, because it is in thermal equilibrium]. If an equivalent
illumination can be realized optically, then the incident power is distributed
equally among the 1-dimensional transverse modes in front of the screen --- as in the
electronic problem. The transmission probability $T$ (which appears in the
Landauer formula (\ref{eq11}), see section \ref{sec6}) is defined as the ratio of the total
transmitted power $P_{\rm trans}$ to the incident power per mode (both per unit of length in the $z$-direction). The latter quantity is $\lambda/2$ times the total incident energy flux $j_{\rm in}$,
so that
\be
T \equiv\frac{2P_{\mathrm{t}\mathrm{r}\mathrm{a}\mathrm{n}\mathrm{s}}}{j_{\mathrm{i}\mathrm{n}}\lambda}.  \label{eq6}
\ee
The one-to-one correspondence with the electron transport problem now tells us
that $T$ is approximately quantized to the number $N$ of $1\mathrm{D}$ transverse modes in the slit,
\be
T\approx N=\mathrm{I}\mathrm{n}\mathrm{t}\left[\frac{2W}{\lambda}\right].   \label{eq7}
\ee
On increasing the width $W$ of the slit one would thus see a {\it step} {\it wise\/} increase of the
transmitted power, if the incident flux is kept constant. The height of the steps is
$j_{\mathrm{in}}\lambda/2$. To observe well-developed plateaus the screen should have a certain
thickness $d$ in order to suppress the evanescent waves through the slit. A thickness
$d\approx (W\lambda)^{1/2}$ is sufficient (see Ref.\ \cite{ref22}). If $d$ is much greater, the transmission steps can become obscured by Fabry-Perot like resonances from waves reflected at the 
front and back end of the slit. These can be avoided by smoothing the edges of the
slit. Additional interference structure (not present in the electronic case) can occur
if light incident from different angles is coherent.

Instead of the polarization given above, one can also use the polarization $B_{\mathrm{x}}=B_{\mathrm{y}}=E_{\mathrm{z}}=0$. The wave function $\Psi$ then corresponds to the magnetic field
component $B_{\mathrm{z}}$. The boundary condition on $B_{\mathrm{z}}$ is that its normal derivative
vanishes on the screen. Although it is not obvious how to realize this boundary
condition in the electronic case, the transmission steps are probably present
irrespective of the boundary conditions (the conductance quantization in a 2DEG
occurs both for smooth and steep potential walls).

The above geometry was chosen to achieve a mapping onto the scalar
two-dimensional electron transport problem. In $3\mathrm{D}$ an electronic system showing
the conductance quantization has not yet been realized experimentally, although it
should be possible in principle. The $3\mathrm{D}$ optical analogue may be more readily
realizable. Consider a metallic screen in the $x-y$ plane with an aperture of arbitrary
shape. At a frequency $\omega$ a number $N$ of 2-dimensional transverse modes in the
aperture are below cut-off. We would expect a {\it step} {\it wise\/} increase in the
transmitted power on increasing the size of the aperture (i.e.\ on increasing $N$), if
the aperture is illuminated in such a way that the incident flux is distributed
equally among the $2\mathrm{D}$ transverse modes in front of the screen. The number $dN$ of
these modes per unit area with wave vectors within a solid angle $d\Omega$ is (including
both polarizations)
\be
dn=2\frac{dk_{\mathrm{y}}}{2\pi}\frac{dk_{\mathrm{z}}}{2\pi}=\frac{k^{2}}{2\pi^{2}}\cos\theta\, d\Omega,   \label{eq8}
\ee
where $\theta$ is the angle with the normal to the screen. It follows that (as in the $2\mathrm{D}$
case discussed earlier) a $\cos\theta$ distribution of the incident flux has the required
equipartition of power among the transverse modes. This angular distribution is
realized e.g.\ by the light scattered diffusely from a surface in front of the screen.
The height of the transmission steps should be the incident power per mode. Now,
the total number $n$ of $2\mathrm{D}$ transverse modes per unit area is
\be
n=\int_{0}^{2\pi}d\phi\int_{0}^{\pi/2} \frac{dn}{d\Omega}\sin\theta d\theta=\frac{2\pi}{\lambda^{2}}. \label{eq9}
\ee
We therefore predict for the $3\mathrm{D}$ case that the height of the steps in the transmitted
power (for an energy flux $j_{\mathrm{i}\mathrm{n}}$ incident on the screen) will be $j_{\mathrm{i}\mathrm{n}}\lambda^{2}/2\pi$, assuming
that the two independent polarizations of the modes in the aperture can be
resolved (the steps are a factor of two larger if this is not the case).

It would be interesting to carry out this analogy between optics and
micro-electronics experimentally.

\section{\label{sec5} Classical electron focusing}

\begin{figure}
\centerline{\includegraphics[width=8cm]{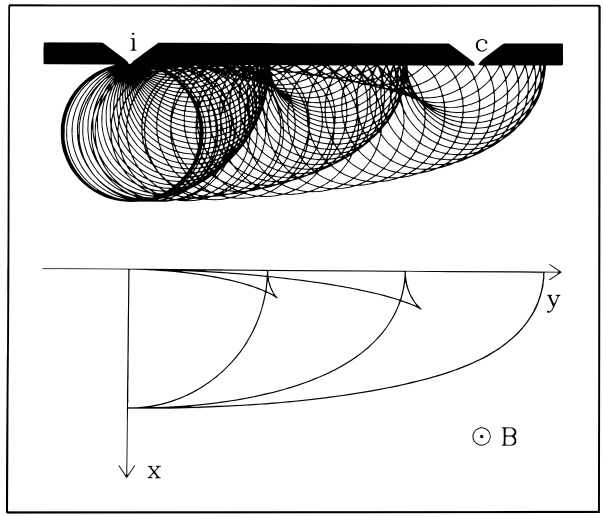}}
\caption{
Skipping orbits at a $2\mathrm{D}\mathrm{E}\mathrm{G}$ boundary. The gate defining the injector (i)
and collector (c) point contacts and the boundary is shown
schematically in black. For clarity the trajectories are drawn up to the
third specular reflection only. Bottom: Calculated location of the caustic
curves. [From Ref.\cite{ref10}]
\label{fig2}
}
\end{figure}

Ballistic transport is in essence a transmission problem. One example is the
conductance of a quantum point contact, which is determined by the total
transmission probability through the constriction, in analogy with the transmission
through a waveguide ({\it cf.\/} secs.\ \ref{sec3} and \ref{sec4}). More sophisticated transmission
experiments can be realized if one point contact is used as a collector. Electron
focusing in the transverse field geometry due to Tsoi,\cite{ref4} is one of the simplest
realizations of such an experiment. (Longitudinal electron focusing\cite{ref3} is not
possible in $2\mathrm{D}$). In our experiments, two adjacent point contacts are positioned on
the same $2\mathrm{D}\mathrm{E}\mathrm{G}$ boundary. The boundary as well as both point contacts are defined
electrostatically in the $2\mathrm{D}\mathrm{E}\mathrm{G}$ by means of a gate electrode of suitable shape, see
Fig.\ \ref{fig2}. A constant current $I_{\mathrm{i}}$ is injected through one of the point contacts, and the
voltage $V_{\mathrm{c}}$ on the second point contact (the collector) is measured. A transverse
magnetic field is used to deflect the injected electrons, such that they propagate in
{\em skipping orbits\/} along the $2\mathrm{D}\mathrm{E}\mathrm{G}$ boundary from the injector towards the collector
--- provided the boundary scattering is specular. In Fig.\ \ref{fig2} the skipping orbits are
illustrated. A magnetic field acts as a {\it lens}, focusing the injected electron beam, as
is evidenced in Fig.\ \ref{fig2} by the presence of {\it caustics\/} or lines of focus (here the classical
density of injected electrons is infinite). The focal points on the $2\mathrm{D}\mathrm{E}\mathrm{G}$ boundary
are separated by the classical cyclotron orbit diameter $2l_{\mathrm{c}\mathrm{y}\mathrm{c}1}=2mv_{\mathrm{F}}/eB$. The classical transmission probability from injector to collector has a peak if a focus
coincides with the collector, which happens if the distance between the two point
contacts $L=p\times 2l_{\rm cycl}$, with $p=1,2,\ldots$. As a result, the collector voltage as a function of magnetic field shows a series of peaks, called a focusing spectrum.

\begin{figure}
\centerline{\includegraphics[width=8cm]{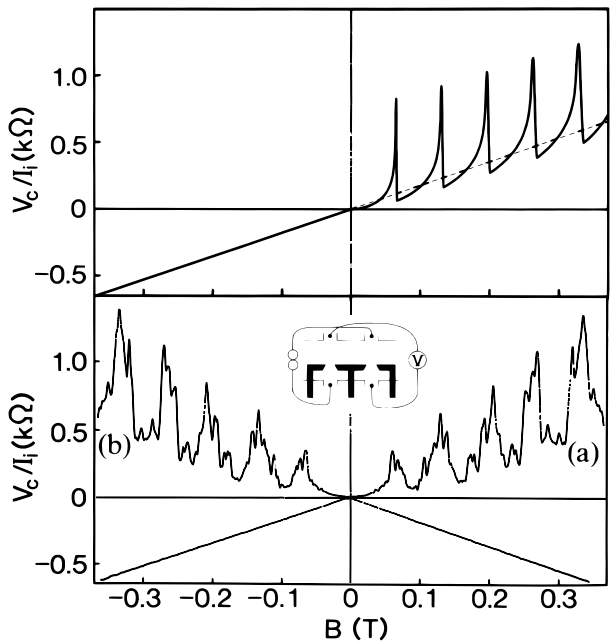}}
\caption{
Top: Classical focusing spectrum, calculated with $W_{\mathrm{i}}=W_{\mathrm{c}}=50\,\mathrm{n}\mathrm{m}$, and
$L=3.0\,\mu \mathrm{m}$. The dashed line is the extrapolation of the classical Hall
resistance seen in reverse fields. Bottom: Experimental electron focusing
spectra obtained in the measurement configuration illustrated in the
inset. The two traces a and b have been obtained by interchanging current and
voltage leads, and demonstrate the injector-collector reciprocity of (\ref{eq13}). [From Ref.\cite{ref9}]
\label{fig3}
}
\end{figure}

The classical transmission probability can be calculated straightforwardly,\cite{ref9}
as in the $3\mathrm{D}$ metal case.\cite{ref28} A plot of the classical focusing spectrum assuming
purely specular boundary scattering is given in Fig.\ \ref{fig3} (top panel) using the experimental
parameters $L=3.0\,\mu \mathrm{m}$ and $k_{\mathrm{F}}=1.5\cdot10^{8}\,\mathrm{m}^{-1}$ and an estimated injector and
collector width of 50 $\mathrm{n}\mathrm{m}$. The spectrum consists of a series of equidistant peaks of
constant amplitude at magnetic fields which are multiples of $B_{\mathrm{f}\mathrm{o}\mathrm{c}\mathrm{u}\mathrm{s}}\equiv 2\hbar k_{\mathrm{F}}/eL\approx
0.066\,{\rm T}$. The focusing signal oscillates around a value given by the conventional
Hall resistance $B/n_{\mathrm{s}}e$ seen in the reverse field signal. In reverse fields no focusing
peaks occur, because the electrons are deflected away from the collector. Note also
that $V_{\mathrm{c}}/I_{\mathrm{i}}$ is approximately quadratic in $B$ for weak positive fields in contrast to
the linear Hall resistance for reverse fields. These features are strikingly confirmed
by the experiment (see Fig.\ \ref{fig3}, bottom panel), which constitutes a direct observation of skipping
orbits at the $2\mathrm{D}\mathrm{E}\mathrm{G}$ boundary, and demonstrates that boundary scattering is highly
specular (because the peaks at higher magnetic fields, corresponding to electrons multiply scattered by the boundary, do not diminish in amplitude). In contrast, in metals the amplitude of subsequent peaks usually diminishes rapidly, indicating partially diffuse boundary scattering.\cite{ref4,ref28} Specular scattering occurs if the Fermi
wavelength is large compared to the spatial scale of the boundary roughness. This
is difficult to achieve in metals (where $\lambda_{\mathrm{F}}\approx 0.5\,\mathrm{n}\mathrm{m}$), but not in a $2\mathrm{D}\mathrm{E}\mathrm{G}$ (where $\lambda_{\mathrm{F}}$
is 100 times larger). This explains why specular scattering is found to be
predominant in our experiments.\cite{ref8,ref9}

While the overall shape of the focusing spectra is as one expects from the
classical calculation, an additional unexpected oscillatory structure is observed in
the experiment. This is the signature of a new phenomenon: coherent electron
focusing.\cite{ref8} As discussed in \ref{sec7}, its origin is mode interference, the relevant
modes being magnetic edge states coherently excited by the injecting point
contact. Fig.\ \ref{fig3} (lower panel) also shows the collector signal obtained after interchanging
current and voltage leads, demonstrating the reproducibility of the fine structure.
The symmetry of the focusing spectrum on interchanging injector and collector is
an example of the Onsager-Casimir relation for the {\em conductance\/} (as opposed to conductivity) derived for quantum ballistic transport by B\"{u}ttiker\cite{ref29}
(see sec.\ \ref{sec6}). A related source-detector {\it reciprocity} {\it theorem\/} is well known in
microwave transmission theory.\cite{ref30} In optics this is known as the reciprocity
theorem of Helmholtz.\cite{ref27}

\section{\label{sec6} Electron focusing as a transmission problem}

\begin{figure}
\centerline{\includegraphics[width=6cm]{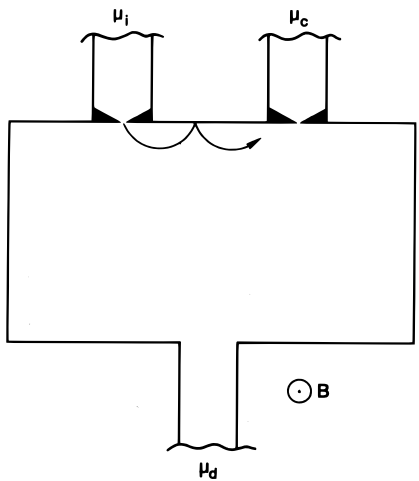}}
\caption{
Three-terminal conductor in the electron focusing geometry. Three
reservoirs at chemical potentials $\mu_{\mathrm{i}},$ $\mu_{\mathrm{c}}$, and $\mu_{\mathrm{d}}$ are connected by leads to
a wide $2\mathrm{D}\mathrm{E}\mathrm{G}$. Two of the leads contain a narrow constriction, or point
contact (shown in black). The current flows from reservoir i to reservoir
d, while reservoir c draws no net current. [From Ref.\cite{ref9}]
\label{fig4}
}
\end{figure}

Transport in the regime of diffusive motion is usually treated in terms of a local
distribution function found from a self consistent solution of the linearized
Boltzmann equation and the Poisson equation. Ballistic transport is inherently
non-local, and can more naturally be viewed as a transmission problem. In the
linear transport regime there is then no need to consider the self consistent electric
field explicitly, which greatly simplifies the analysis. This approach originated in a
paper by Landauer in 1957,\cite{ref31} and has since been generalized and extended by
B\"{u}ttiker\cite{ref29} to the case of a realistic conductor with multiple leads connected to
the external current source and voltmeters. The equivalence of B\"{u}ttikers approach
and the more familiar linear response theory based on the Kubo formalism has
been demonstrated.\cite{ref32,ref33,ref34} In this section we apply the Landauer-B\"{u}ttiker
formula to the electron focusing geometry. For simplicity, we consider the three-terminal configuration of Fig.\ \ref{fig4}, with point contacts in two of the probes,
serving as injector and collector. (By a simple extension of the arguments in this
section, one can also derive expressions for the four-terminal geometry of Fig.\ \ref{fig3}.\cite{ref9} In order to have well defined initial and final states for the scattering
problem, the probes are connected via idealized leads (or electron waveguides) to
reservoirs at a constant electro-chemical potential. We denote by $\mu_{\mathrm{i}}$ and $\mu_{\mathrm{c}}$ the
chemical potentials of the injector and collector reservoirs. The chemical potential
of the drain reservoir, which acts as a current sink, is assigned the value zero.

Following B\"{u}ttiker,\cite{ref29} we can relate the currents $I_{\alpha}$ $(\alpha=\rm{i,c,d})$ in the leads to
these chemical potentials via the transmission probabilities $T_{\alpha\rightarrow\beta}$ from reservoir $\alpha$
to $\beta$, and reflection probabilities $R_{\alpha}$ (from reservoir $\alpha$ back to the same reservoir).
These equations have the form
\be
\frac{h}{2e}I_{\alpha}=(N_{\alpha}-R_{\alpha})\mu_{\alpha}-\sum_{\beta\neq\alpha}T_{\beta\rightarrow\alpha}\mu_{\beta}.\label{eq10}
\ee
Here $N_{\alpha}$ is the number of occupied (spin degenerate) transverse waveguide modes
or ``channels'' in lead $\alpha$. Eq.\ (\ref{eq10}) can also be used in the classical limit, where the
discreteness of $N_{\alpha}$ can be ignored. We first apply these equations to the
two-terminal conductance of a single point contact. In that case all $\mu_{\beta}=0$ for
$\beta\neq\alpha$, and thus
\be
G\equiv\frac{I_{\alpha}}{\mu_{\alpha}/e}=\frac{2e^{2}}{h}T,   \label{eq11}
\ee
with $T\equiv N_{\alpha}-R_{\alpha}$ the total transmission probability for the $N_{\alpha}$ channels in the
leads. As discussed in sec.\ \ref{sec3}, for point contacts with $W\geq\lambda_{\mathrm{F}}$ one has in a good
approximation $T\approx N$, with $N\approx k_{\mathrm{F}}W/\pi$ the number of occupied subbands in the
constriction.

In the electron focusing experiments the collector is connected to a voltmeter,
which implies $I_{\mathrm{c}}=0$ and $I_{\mathrm{d}}=-I_{\mathrm{i}}$. We then find from (\ref{eq10})
\begin{subequations}
\label{eq12}
\begin{eqnarray}
&&\mu_{\mathrm{c}}=\frac{T_{\mathrm{i}\rightarrow \mathrm{c}}}{N_{\mathrm{c}}-R_{\mathrm{c}}}\mu_{\mathrm{i}},\label{eq12a}\\
&&\frac{h}{2e}I_{\mathrm{i}}=(N_{\mathrm{i}}-R_{\mathrm{i}})\mu_{\mathrm{i}}-T_{\mathrm{c}\rightarrow \mathrm{i}}\mu_{\mathrm{c}}. \label{eq12b}
\end{eqnarray}
\end{subequations}
The transmission probability $T_{\mathrm{c}\rightarrow \mathrm{i}}\approx 0$ because the magnetic field deflects
electrons from the injector towards the drain, and away from the collector (see Fig.\ \ref{fig4}).
The measured quantity is the ratio of collector voltage $V_{\mathrm{c}}\equiv\mu_{\mathrm{c}}/e$ (relative to
the voltage of the drain) and injector current
\be
\frac{V_{\mathrm{c}}}{I_{\mathrm{i}}}=\frac{2e^{2}}{h}\frac{T_{\mathrm{i}\mathrm{c}}}{G_{\mathrm{i}}G_{\mathrm{c}}}, \label{eq13}
\ee
where we have used (\ref{eq11}). The transmission probability $T_{\rm{i}\rightarrow{\rm c}}$ is evaluated in the next sections. This result is symmetric under interchange of injector and collector leads, with simultaneous reversal of the magnetic field, because\cite{ref29} $T_{\rm{i}\rightarrow{\rm c}}(B)=T_{\rm{c}\rightarrow\rm{i}}(-B)$. This symmetry relation explains the injector-collector reciprocity observed experimentally (see Fig.\ \ref{fig3}, traces a and b).

\section{\label{sec7} Coherent electron focusing}

\subsection{\label{sec7.1} Experiment}

\begin{figure}
\centerline{\includegraphics[width=8cm]{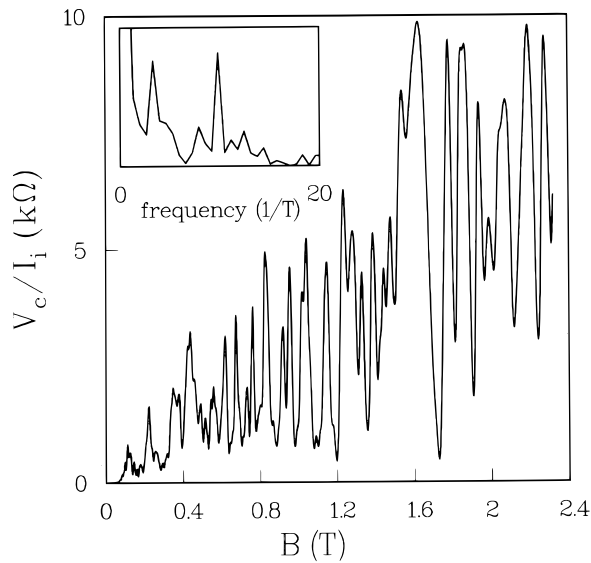}}
\caption{
Electron focusing spectra measured at 50 mK for a device with 1.5 $\mu{\rm m}$ point contact separation, showing large quantum interference structure. The inset gives the Fourier transform power spectrum, for $B>0.8\,{\rm T}$, demonstrating that the dominant periodicity is the low field classical focusing periodicity. [From Ref.\cite{ref9}]
\label{fig5}
}
\end{figure}

The oscillatory structure superimposed on the classical focusing peaks in the
experimental trace of Fig.\ \ref{fig3} is a quantum interference effect. At higher magnetic
fields (beyond about 0.4 $\mathrm{T}$), the collector voltage shows oscillations with a much
larger amplitude than the low field focusing peaks, and the resemblance to the
classical focusing spectrum is lost. The oscillatory structure becomes especially
dramatic on decreasing the width of the point contacts (by increasing the gate
voltage), and at very low temperatures, if also the voltage drop across the injector
point contact is maintained below $k_{\mathrm{B}}T/e$. In Fig.\ \ref{fig5} we show the experimental
results for a device with 1.5 $\mu \mathrm{m}$ point contact separation, and estimated point
contact width of $20$--$40$ nm. Note the nearly 100\% modulation of the focusing
signal. A Fourier transform of the data (see inset of Fig.\ \ref{fig5}) shows that the
large-amplitude high-field oscillations have a dominant periodicity of 100 mT,
approximately the same as the periodicity $B_{\mathrm{f}\mathrm{o}\mathrm{c}\mathrm{u}\mathrm{s}}$ of the low field focusing peaks.
This dominant periodicity is insensitive to changes in the gate voltage, and is the
characteristic feature of coherent electron focusing which is most amenable to a
direct comparison with theory.

In this section we present a theory of coherent electron focusing.\cite{ref9,ref10} We start
with a short discussion of magnetic edge states, which are the modes of this
transmission problem. After a brief discussion of the excitation of these modes by
the injecting point contact, we proceed by giving an explanation of the characteristic features of the observed spectrum in terms of {\em mode interference}. This treatment was first given in Ref.\ \cite{ref10}, and in a more detailed form in Ref.\ \cite{ref9}, together with an equivalent one in the ray-picture (interference between trajectories).

\subsection{\label{sec7.2} Skipping orbits and magnetic edge states}

\begin{figure}
\centerline{\includegraphics[width=8cm]{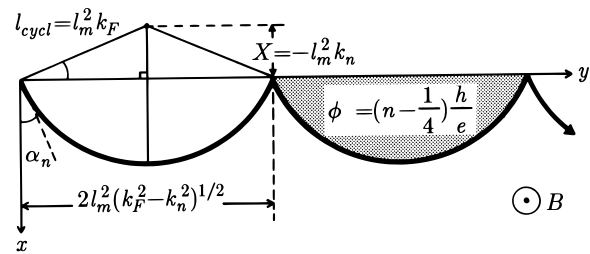}}
\caption{
Skipping orbit at a 2DEG boundary. The semi-classical correspondence with a magnetic edge state is indicated. The flux $\phi$ enclosed in the shaded area equals $(n-\frac{1}{4})$ elementary flux quanta $h/e$.
\label{fig6}
}
\end{figure}

The motion of conduction electrons at the boundaries of a 2DEG in a perpendicular magnetic field is in skipping orbits, provided the boundary scattering is specular (see Figs.\ \ref{fig2} and \ref{fig6}). (In a narrow channel traversing states can coexist with skipping orbits, as discussed {\em e.g.\/} in Ref.\ \cite{ref16}). The position ($x,y$) of the electron on the circle with center coordinates ($X,Y$) can be expressed in terms of its velocity by
\be
x=X+v_{\rm y}/\omega_{\rm c},\;\;y=Y-v_{\rm x}/\omega_{\rm c},\label{eq14}
\ee
with $\omega_{\mathrm{c}}=eB/m$ the cyclotron frequency. The separation $X$ of the center from the
boundary is constant on a skipping orbit, while $Y$ jumps over a distance equal to
the chord length on each specular reflection. The canonical momentum $\bm{p}=m\bm{v}-e\bm{A}$ is given in the Landau gauge $\bm{A}=(0,Bx,0)$ by
\be
p_{\mathrm{x}}=mv_{\mathrm{x}},\;\; p_{\mathrm{y}}=-eBX,   \label{eq15}
\ee
so that $p_{\mathrm{y}}$ is a constant of the motion. The periodic motion perpendicular to the
boundary leads to the formation of discrete quantized states (with quantum
number $n$). These states are known as magnetic edge states (and in metals as
magnetic surface states\cite{ref35,ref36}). The wave number {\it along\/} the boundary is a plane
wave with wave number $k_{\rm y}\equiv p_{\rm y}/\hbar$, which can be expressed in the guiding center coordinate $X$ according to $X=-l_{\rm m}^{2}k_{\rm y}$. Here $l_{\rm m}$ is the magnetic
length, which plays a role similar to that of $\lambda_{\mathrm{F}}$ in the absence of a magnetic field.
Coherent electron focusing constitutes an interference experiment with quantum
ege states at the Fermi energy. The wave number $k_{\rm y}$ for thes states has a quantized value, denoted by $k_{\rm n}$. The phase of each edge state arriving at the
collector is determined by $k_{\mathrm{n}}$, and the functional dependence of $k_{\mathrm{n}}$ on $n$ is thus
important.

An exact quantum mechanical treatment is possible, but here it suffices to
consider the semi-classical Bohr-Sommerfeld approximation applied to the
periodic motion in the $x$ direction
\be
\frac{1}{\hbar}\int p_{\mathrm{x}}\,dx+\gamma=2\pi n.   \label{eq16}
\ee
The integral is over one period of the motion, and $\gamma$ is the sum of the phase shifts
acquired at the two classical turning points.\footnote{
An essential difference between light optics and electron optics is hidden in this
expression. Wave fronts are perpendicular to light rays. This is not the case for
electrons in a magnetic field, because the phase velocity (in the direction of the
canonical momentum $\bm{p}$) is not parallel to the electron velocity $\bm{v}$. Due to this ``skew connection\cite{ref27}'' the phase difference between paths connecting two points is
determined by the sum of the optical path length difference and the
Aharonov-Bohm phase coresponding to the magnetic flux piercing the area
enclosed. This phase is included via the canonical momentum in (\ref{eq16}), and has to be
taken into account explicitly in a trajectory treatment of coherent electron
focusing.\cite{ref9,ref10}
}
The phase shift upon reflection at the
boundary is $\pi$, as in optics upon reflection of a metallic mirror. The other turning
point is a {\it caustic}, giving rise to a phase shift\footnote{
A lens cannot focus a particle flux tube to a point, because diffraction sets a lower
limit to the flux tube cross section.\cite{ref37} At a small separation $R$ from a caustic the cross section is proportional to $R$, so that the amplitude $A\propto R^{-1/2}$. The sign change of $R$ upon passing through the caustic then leads to the phase factor $(-1)^{-1/2}=\exp(-i\pi/2)$.
}
of $-\pi/2$. Using (\ref{eq14},\ref{eq15},\ref{eq16}) we can thus write
\be
\frac{eB}{m}\int(Y-y)\,dx=2\pi(n-\tfrac{1}{4}),\;\;n=1,2,\ldots   \label{eq17}
\ee
This quantization rule has the simple geometrical interpretation that the flux
enclosed by one arc of the skipping orbit and the boundary equals $(n-\frac{1}{4})$ times the
flux quantum $h/e$ (see Fig.\ \ref{fig6}). This implies that the angle $\alpha_{\mathrm{n}}$, under which the
skipping orbit is reflected from the boundary, is quantized. Simple geometry shows that
\be
\frac{\pi}{2}-\alpha_{\mathrm{n}}-\tfrac{1}{2}\sin 2 \alpha_{\mathrm{n}}=\frac{2\pi}{k_{\mathrm{F}}l_{\mathrm{c}\mathrm{y}\mathrm{c}1}}(n-\tfrac{1}{4}),\;\;n=1,2,\ldots N, \label{eq18}
\ee
with $N$ the largest integer smaller than $\frac{1}{2}k_{\rm F}l_{\rm cycl}+\frac{1}{4}$. (For simplicity, we approximate $N\approx\frac{1}{2}k_{\rm F}l_{\rm cycl}$ in the rest of this paper.) It follows from Fig.\ \ref{fig6} that $k_{n}=k_{\rm F}\sin\alpha_{n}$. The dependence of the wavenumber of each edge state on the mode index
$n$ is thus implicitly contained in (\ref{eq18}).

\subsection{\label{sec7.3} Mode-interference and coherent electron focusing}

In optics, a coherent point source results if a small hole in a screen is illuminated
by an extended source. Such a point source excites the states on the other side of
the screen coherently. A small point contact acts similarly, at least in weak
magnetic fields, which is the reason why we can perform coherent electron focusing
experiments. The wave function $\Psi$ resulting from the coherent excitation at $y=0$
is of the form
\be
\Psi(x,y)=\sum_{n=1}^{N}w_{\rm n}f_{\mathrm{n}}(x)\exp(ik_{\mathrm{n}}y).   \label{eq19}
\ee
Here $f_{n}(x)$ is the transverse wave function of mode $n$, and $w_{n}$ its excitation factor. The collector voltage is in a first approximation (in the regime $W\ll\lambda_{\rm F}$), determined by the probability density of the wave function at an infinitesimal distance from the boundary, unperturbed by the presence of the collector point contact. In the coherent focusing regime, the probability density near the collector at $y=L$ is dominated by the phase factors $\exp(ik_{n}L)$, which vary rapidly as a function of $n$. The weight factors $w_{n}$ depend on the modeling of the point contacts, but do not affect the qualitative features of the focusing spectrum.

For point contacts modeled as an ideal point source, the transmission
probability from injector to collector $T_{\mathrm{i}\rightarrow\mathrm{c}}$ occurring in the general formula (\ref{eq13})
can be written as a product of three sequential transmission probabilities:\footnote{
This is in contrast to the opposite limit of adiabatic transport, realized for wider
point contacts in strong magnetic fields.\cite{ref9,ref38}
}
\begin{enumerate}
\item Through the injector point contact with probability $G_{\rm i}/(2e^{2}/h)$.
\item From the vicinity of the injector at $(x,y)=(0,0)$ to a point $(0,L)$ near the collector
point contact. This is directly proportional to the probability density
close to the collector, which can be expressed as the square of a sum of $N$ plane waves ({\em cf.} (\ref{eq19})).
\item From there through the collector with probability $G_{\rm c}/(2e^{2}/h)$. 
\end{enumerate}
Using a weight factor derived in Ref.\ \cite{ref9}, we thus find
\begin{subequations}
\label{eq20}
\begin{eqnarray}
\frac{V_{\rm c}}{I_{\mathrm{i}}}&=& \frac{h}{2e^{2}}\left|\frac{1}{N}\sum_{n=1}^{N}\mathrm{e}^{ik_{n}L}\right|^{2}\label{eq20a}\\
&=&\frac{h}{2e^{2}}\frac{1}{N^{2}}\left[N+\sum_{n\neq n'}\sum_{n'}\cos[(k_{n}-k_{n'})L]\right].\nonumber\\
&& \label{eq20b}
\end{eqnarray}
\end{subequations}
The first term in (\ref{eq20b}) is the incoherent contribution, which gives $V_{\mathrm{c}}/I_{\mathrm{i}}=(h/2e^{2})N^{-1}$. This is just the ordinary Hall effect, without corrections due to focusing. The second term in (\ref{eq20b}) represents a sum of oscillations in the
transmission probability, expressed in terms of interference between pairs of edge
states with different wave numbers.

\begin{figure}
\centerline{\includegraphics[width=8cm]{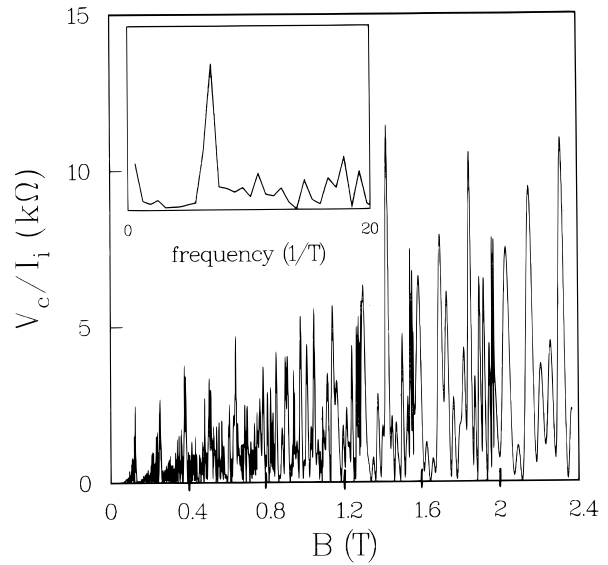}}
\caption{
Theoretical electron focusing spectrum calculated from (\ref{eq20}) for $L=1.5$
$\mu \mathrm{m}$. Inset shows the Fourier power spectrum for $B>0.8$ T. No
correction has been made for the finite width of the point contacts in the
experiment of Fig.\ \ref{fig5}.
\label{fig7}
}
\end{figure}

In the experiment $B$ is varied, affecting the values of $k_{\mathrm{n}}$ and $N$. A plot of a
numerical evaluation of (\ref{eq20}) (for the experimental parameters of Fig.\ \ref{fig5}) is shown in
Fig.\ \ref{fig7}. As in the experiment, we find fine structure\footnote{
We have not attempted to model the partial spatial coherence due to the finite
point contact size. This would introduce a coherence length into the problem,
reflecting the properties of the point contact source, to be distinguished from the
coherence length related to inelastic scattering, known in the field of transport in
disordered conductors,\cite{ref12,ref13,ref18} which is a property of the medium.
}
on classical focusing peaks at
low magnetic fields, which becomes entirely dominant at higher fields. It is
apparent from Fig.\ \ref{fig7} (and confirmed by Fourier transform) that the
{\it large-amplitude} {\it high-field} {\it oscillations} {\it have} {\it the} {\it same} {\it periodicity} {\it as} {\it the} {\it smaller low-field peaks} --- as observed experimentally. This is the main result of our
calculation which we have found to be insensitive to details of the point contact
modeling. (We have checked numerically that contributions due to evanescent
waves, neglected in (\ref{eq20}), are small). The calculation presented here assumes that
all edge states are excited by the point contact (see Ref.\ \cite{ref9}). The case of selective
excitation of edge states which results if the point contacts are flared into a horn is
discussed elsewhere.\cite{ref11}

If all $N$ modes arrive in phase at the collector, (\ref{eq20}) yields the {\em upper bound\/} of the focusing signal $V_{\rm c}/I_{\rm i}=h/2e^{2}\approx 12.9\,{\rm k}\Omega$, enhanced by a factor $N$ over the incoherent result. The minimum intensity due to destructive interference is zero. The largest
oscillations observed experimentally (see Fig.\ \ref{fig5}) are from 0.3 to 10 $\mathrm{k}\Omega$, which is
close to these limits. Peaks of comparable magnitude are seen in the calculated
spectrum (Fig.\ \ref{fig7}). This demonstrates that a nearly ideal coherence between the
different edge states has been realized in this experiment.

\begin{figure}
\centerline{\includegraphics[width=8cm]{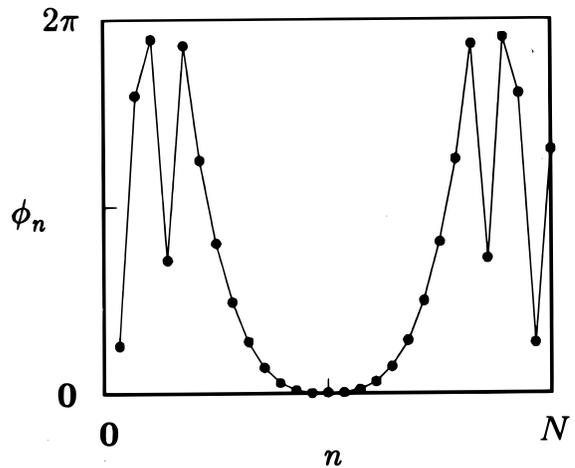}}
\caption{
Phase $\phi_{\mathrm{n}}=k_{\mathrm{n}}L$ (modulo $2\pi$) of each edge state at the collector for
$L/2l_{\mathrm{cycl}}=2$, and $N=k_{\rm F}l_{\rm cycl}/2=28$. Edge states with $n$ in the neighborhood of $N/2$ interfere constructively.
\label{fig8}
}
\end{figure}

We now turn to a qualitative discussion of the origin of the periodicity in the
focusing spectrum. The dependence on $n$ of the phase $k_{\mathrm{n}}L$ at the collector is close
to linear in a broad interval. Expansion of (\ref{eq18}) around $\alpha_{\mathrm{n}}=0$ gives
\be
k_{\mathrm{n}}L={\rm constant} - 2 \pi\frac{B}{B_{\mathrm{f}\mathrm{o}\mathrm{c}\mathrm{u}\mathrm{s}}}+k_{\mathrm{F}}L\times{\rm order}\left[\frac{N-2n}{N}\right]^{3}. \label{eq21}
\ee
Here $B_{\mathrm{f}\mathrm{o}\mathrm{c}\mathrm{u}\mathrm{s}}\equiv 2\hbar k_{\mathrm{F}}/eL$ is the same as the magnetic field which follows from the
classical focusing condition $2l_{\mathrm{c}\mathrm{y}\mathrm{cl}}=L$ discussed in Sec.\ \ref{sec5}. It follows from this expansion that, if $B/B_{\rm focus}$ is an integer, a fraction of order $(1/k_{\rm F}L)^{1/3}$ of the $N$ edge states interfere constructively at the collector. Because of the $1/3$ power this is a substantial fraction, even for the large $k_{\mathrm{F}}L\approx 225$ of the experiment. This is
confirmed by the plot in Fig.\ \ref{fig8} of the phase for each edge state (modulo $2\pi$)
obtained from a numerical solution of (\ref{eq18}), as a function of $n$ for the second
focusing peak. We conclude that the enhanced magnitude of the high field peaks
with the classical focusing periodicity is a consequence of the constructive
interference of a large fraction of the coherently excited edge states. The classical
focusing spectrum\footnote{
To be more precise, in the limit $\lambda_{\mathrm{F}}/L\rightarrow 0$, $\lambda_{\mathrm{F}}/W=$ constant, one obtains focusing
peaks with the classical $B_{\mathrm{f}\mathrm{o}\mathrm{c}\mathrm{u}\mathrm{s}}$ periodicity, and with negligible fine structure.
However, the shape and height of the peaks will be different from the classical
result in Fig.\ \ref{fig3} (top panel), if one retains the condition that the point contact width $W\lesssim\lambda_{\mathrm{F}}$.
For a fully classical focusing spectrum one needs also that $\lambda_{\mathrm{F}}/W\rightarrow 0$.
}
is regained in the limit $\lambda_{\mathrm{F}}/L\rightarrow 0$. Raising the temperature
induces a smearing of the focusing spectrum, but the line shape does not approach
the classical one.\cite{ref9}

The edge states outside the domain of linear $n$-dependence of the phase give
rise to additional interference structure, which does not have a simple periodicity.
From (\ref{eq20b}), it is clear that the oscillations are, generally, determined by the {\it mode}
{\it interference} {\it wavelengths}
\be
\Lambda_{n,n'}\equiv 2\pi/(k_{n}-k_{n'}).   \label{eq22}
\ee
Two modes $n$ and $n'$ arrive in phase at the collector, and thus interfere
constructively, if the distance between injector and collector $L$ is an integer
multiple of their mode-interference wavelength. The wave number $k_{\mathrm{n}}$ has values
between $\pm k_{\mathrm{F}}$. It follows from (\ref{eq18}) that the {\it largest\/} mode interference wavelength
$\Lambda_{\rm max}=2l_{\mathrm{c}\mathrm{y}\mathrm{cl}}$ (corresponding to interference between mode $n=N/2$ with its
nearest neighbors\footnote{
Using $k_{n}=k_{\rm F}\sin\alpha_{n}$ and differentiating (\ref{eq18}), we find a general expression for the mode-interference wavelength for neighboring modes $\Lambda_{n,n-1}=2l_{\rm m}^{2}\{k_{\rm F}^{2}-k_{n}^{2}\}^{1/2}$ which is recognized as the chord length of the skipping orbit corresponding to the edge state $n$ (see Fig.\ \ref{fig6}).
}). This wavelength is associated with the dominant periodic
oscillations, discussed above. The {\it shortest\/} mode interference wavelength is $\Lambda_{\rm min}=\pi/k_{\mathrm{F}}$ (corresponding to interference between mode $n=0$ and $n=N$). This is
shorter by a factor $\pi/2k_{\rm F}l_{\rm cycl}\approx 1/N$, the number of edge states. This explains why the fast oscillations in the focusing spectrum disappear at strong fields, where only a few edge states are occupied. We remark that in the latter case the argument
based on (\ref{eq21}) breaks down, and as witnessed by the magnitude of some of the
focusing peaks in Fig.\ \ref{fig7} (or in the experiment), occasionally {\it all\/} the modes interfere
constructively.

Our theory accounts for the most important novel features of the experiment,
which are not observed in the classical ballistic transport regime in metals. A more
quantitative comparison between theory and experiment requires a detailed
analysis of the potential at the point contact and at the $2\mathrm{D}\mathrm{E}\mathrm{G}$ boundary,
something which we have not attempted. Also, we surmise that an exact treatment
of the focusing for infinitesimally narrow point contacts on an exact infinitely
repulsive potential boundary would be possible, since the transmission probabilities
can then be expressed in terms of the unperturbed wave functions of the edge
states --- which are known exactly (Weber functions). Undoubtedly, such a calculation would give results different in detail from the calculated spectrum in Fig.\ \ref{fig7}. It is most likely, however, that --- as in the case of related phenomena mentioned below --- the uncertainties in the experimental conditions\cite{ref9} will preclude a better agreement with an exact theory. We stress that the observed appearance
of high-field oscillations with the focusing periodicity, but with much large
amplitude is characteristic for the mode-interference mechanism proposed.

Coherent electron focusing is a nice demonstration of a transmission experiment
in the quantum ballistic transport regime. It has also yielded information of a more
specific nature:\cite{ref8,ref9,ref10}
\begin{enumerate}
\item Boundary scattering in a $2\mathrm{D}\mathrm{E}\mathrm{G}$ is highly specular,
capable of sustaining descriptions of ballistic transport in terms of skipping orbits
and magnetic edge states. 
\item Electron motion in the $2\mathrm{D}\mathrm{E}\mathrm{G}$ is ballistic and coherent
over distances of several microns. 
\item A small quantum point contact acts as a
monochromatic point source, exciting a coherent superposition of edge states. 
\end{enumerate}
By combining these properties in the electron focusing geometry, a quantum
interference effect leading to a conductance modulation of nearly 100\% is realized.

\section{\label{sec8} Other mode-interference phenomena}

In this paper we have discussed coherent electron focusing as a manifestation of electron optics in the solid state. Such analogies provide us with valuable insight if used with caution, and they can stimulate new experiments. We mention some other interesting analogies taken from different fields. Very long wavelength radio waves ($\lambda$ of the order of 1 km) propagate around the earth as in a waveguide, bounded by the parallel curved conducting surfaces formed by the earth and the ionosphere. The guiding action explains why Marconi in 1902 could be successful in transmitting radio signals across the Atlantic ocean.\cite{ref39} Mode-interference is commonly observed as fading in radio signals at sunrise or sunset. Focusing of guided sound waves occurs in the ocean as a result of a vertical refractive index profile (due to gradients in hydrostatic pressure, salinity and temperature). The mode- and ray-treatments of the resulting interference patterns in the acoustic pressure have many similarities with our theory of coherent electron focusing.\cite{ref40,ref41}. The propagation of curved rays along a plane surface (as in electron focusing), is formally equivalent to that of straight rays along a curved surface. We mention in this connection the clinging of sound to a curved wall, which is the mechanism responsible for the ``whispering gallery'' effect in St.\ Paul's cathedral, explained by Lord Rayleigh,\cite{ref42} and for the talking wall in the Temple of Heaven in Peking.\cite{ref41}

Quantum ballistic transport can be studied in many geometries, and a wealth of results has been obtained by various groups.\cite{ref12,ref13,ref14} Many aspects of the present discussion have an applicability beyond the specific context of electron focusing. For example, the description of transport in terms of excitation, detection and interference of quantum subbands or magnetic edge states as modes in a transmission problem, is equally significant for the Aharonov-Bohm effect in small 2DEG rings or for transport in multiprobe ``electron waveguides''.\cite{ref43,ref44,ref45,ref46,ref47} These concepts have also been applied to the quantum Hall effect.\cite{ref9,ref38,ref48,ref49} Another area of research concerns the reproducible conductance fluctuations observed as a function of magnetic field or gate-voltage in small disordered electronic systems. This field has been extensively explored in the diffusive transport regime, and a standard theory of these ``universal conductance fluctuations'' in terms of quantum interference of the conduction electrons on random paths is available.\cite{ref13,ref50} The physics underlying these fluctuations changes, however, as the ballistic transport regime is approached, because of the increasing importance of boundary scattering.\cite{ref17,ref51} For channels with a width comparable to the Fermi wavelength, a description of transport in terms of subbands, or modes becomes the natural one. A theory for fluctuations in this regime is not yet available. We believe that the concept of mode interference, discussed here for the purely ballistic transport regime, may present a useful point of departure. Some recent theoretical work proceeds in this direction,\cite{ref52,ref53} while several transport measurements may already be in the relevant regime.\cite{ref54}

\acknowledgments

The authors have greatly benefitted from their collaboration with M. E. I. Broekaart, C. T. Foxon, J. J. Harris, L. P. Kouwenhoven, P. H. M. van Loosdrecht, D. van der Marel, J. E. Mooij, J. A. Pals, M. F. H. Schuurmans, B. J. van Wees and J. G. Williamson. Valuable discussions with D. van der Marel on the optical analogue of the conductance quantization are gratefully acknowledged.


\begin{thebibliography}{99}
\bibitem{ref1} O. Klemperer and M. E. Barnett, {\em Electron Optics\/} (Cambridge University Press, Cambridge, 1971).
\bibitem{ref2} H. Busch, Ann.\ Phys.\ Lpz.\ {\bf 81}, 974 (1926).
\bibitem{ref3} Yu. V. Sharvin, Zh. Eksp. Teor. Fiz. \textbf{48}, 984 (1965) [Sov. Phys. JETP 21, 655 (1965)].
\bibitem{ref4} V. S. Tsoi, Pis'ma Zh. Eksp. Teor. Fiz. \textbf{19}, 114 (1974) [JETP Lett. \textbf{19}, 70 (1974)].
\bibitem{ref5} T. Ando, A. B. Fowler, and F. Stern, Rev. Mod. Phys. \textbf{54}, 437 (1982); R. E. Prange and S. M. Girvin, eds., {\em The Quantum Hall Effect\/} (Springer, New York, 1987).
\bibitem{ref6} B. J. van Wees, H. van Houten, C. W. J. Beenakker, J. G. Williamson, J. P. Kouwenhoven, D. van der Marel, and C. T. Foxon, Phys. Rev. Lett. {\bf 60}, 848 (1988); B. J. van Wees {\em et al.}, Phys. Rev. B \textbf{38}, 3625 (1988).
\bibitem{ref7} D. A. Wharam, M. Pepper, H. Ahmed, J. E. F. Frost, D. G. Hasko, D. C. Peacock, D. A.
Ritchie, and G. A. C. Jones, J. Phys. C \textbf{21}, L887 (1988).
\bibitem{ref8} H. van Houten, B. J. van Wees, J. E. Mooij, C. W. J. Beenakker, J. G. Williamson, and
C. T. Foxon, Europhys. Lett. \textbf{5}, 721 (1988).
\bibitem{ref9} H. van Houten, C. W. J. Beenakker, J. G. Williamson, M. E. I. Broekaart, P. H. M. van
Loosdrecht, B. J. van Wees, J. E. Mooij, C. T. Foxon, and J. J. Harris, Phys. Rev. B \textbf{39}, 8556
 (1989).
\bibitem{ref10} C. W. J. Beenakker, H. van Houten, and B. J. van Wees, Europhys. Lett. \textbf{7}, 359 (1988).
\bibitem{ref11} C. W. J. Beenakker, H. van Houten, and B. J. van Wees, Festk\"{o}rperprobleme \textbf{29}, 299 (1989).
\bibitem{ref12} H. Heinrich, G. Bauer, and F. Kuchar, eds., {\em Physics and Technology of Submicron
Structures\/} (Springer, Berlin, 1988).
\bibitem{ref13} P. A. Lee, R. A. Webb and B. L. Al'tshuler, eds., {\em Mesoscopic Phenomena in Solids\/} (Elsevier, Amsterdam, to be published).
\bibitem{ref14} M. Reed and W. P. Kirk, eds., {\em Nanostructure Physics and Fabrication\/} (Academic, New York, 1989).
\bibitem{ref15} K. von Klitzing, G. Dorda, and M. Pepper, Phys. Rev. Lett. \textbf{45}, 494 (1980).
\bibitem{ref16} C. W. J. Beenakker, H. van Houten, and B. J. van Wees, Superlattices and Microstructures \textbf{5}, 127 (1989).
\bibitem{ref17} H. van Houten, B. J. van Wees, and C. W. J. Beenakker, in Ref.\ \cite{ref12}.
\bibitem{ref18} Y. Imry, in {\em Directions in Condensed Matter Physics}, Vol.\ 1, G. Grinstein and G.
Mazenko, eds. (World Scientific, Singapore, 1986).
\bibitem{ref19} L. I. Glazman, G. B. Lesovick, D. E. Khmel'nitskii, R. I. Shekhter, Pis'ma Zh. Teor.
Fiz. \textbf{48}, 218 (1988) [JETP Lett. \textbf{48}, 238 (1988)].
\bibitem{ref20} R. Landauer, Z. Phys. B \textbf{68}, 217 (1987).
\bibitem{ref21} C. W. J. Beenakker and H. van Houten, Phys. Rev. B \textbf{39}, 10445 (1989); H. van Houten and C. W. J. Beenakker, in Ref.\ \cite{ref14}.
\bibitem{ref22} E. G. Haanappel and D. van der Marel, Phys. Rev. B \textbf{39}, 5484 (1989); D. van der Marel and E. G. Haanappel, Phys. Rev. B \textbf{39}, 7811 (1989).
\bibitem{ref23} A. Szafer and A. D. Stone, Phys. Rev. Lett. \textbf{62}, 300 (1989).
\bibitem{ref24} G. Kirczenow, Solid State Comm. \textbf{71}, 469 (1989).
\bibitem{ref25} I. B. Levinson, Pis'ma Zh. Eksp. Teor. Fiz. \textbf{48}, 273 (1988) [JETP Lett. \textbf{48}, 301 (1988)].
\bibitem{ref26} Song He and S. Das Sarma, Phys. Rev. B \textbf{40}, 3379 (1989); E. Tekman and S. Ciraci, Phys. Rev. B \textbf{39}, 8772 (1989); A. Kawabata, J. Phys. Soc. Japan \textbf{58}, 372 (1989).
\bibitem{ref27} M. Born and E. Wolf, {\em Principles of Optics}, 3d rev.\ (Pergamon, Oxford, 1965).
\bibitem{ref28} P. A. M. Benistant, Ph.D. thesis, University of Nijmegen, The Netherlands, 1984; P. A. M. Benistant, A. P. van Gelder, H. van Kempen, and P. Wyder, Phys. Rev. B \textbf{32}, 3351 (1985).
\bibitem{ref29} M. B\"{u}ttiker, Phys. Rev. Lett. \textbf{57}, 1761 (1986); IBM J. Res. Dev. \textbf{32}, 317 (1988).
\bibitem{ref30} G. C. Southworth, {\em Principles and Applications of Waveguide Transmission\/} (Van Nostrand, Toronto, 1950): p.\ 25.
\bibitem{ref31} R. Landauer, IBM J. Res. Dev. \textbf{1}, 223 (1957); see also Ref.\ \cite{ref20}.
\bibitem{ref32} D. S. Fisher and P. A. Lee, Phys. Rev. B \textbf{23}, 6851 (1981).
\bibitem{ref33} A. D. Stone and A. Szafer, IBM J. Res. Dev. \textbf{32}, 384 (1988); H. U. Baranger and A. D. Stone, Phys. Rev. Lett. \textbf{63}, 414 (1989).
\bibitem{ref34} J. Kucera and P. Streda. J. Phys. C \textbf{21}, 4357 (1988).
\bibitem{ref35} R. E. Prange and T.-W. Nee, Phys. Rev. \textbf{168}, 779 (1968).
\bibitem{ref36} M. S. Khaikin, Adv. Phys. \textbf{18}, 1 (1969).
\bibitem{ref37} L. D. Landau and E. M. Lifshitz, \emph{The Classical Theory of Fields}, 4th ed.\ (Pergamon, Oxford, 1987): par.\ 54.
\bibitem{ref38} B. J. van Wees, E. M. M. Willems, C. J. P. M. Harmans, C. W. J. Beenakker, H. van Houten, J.
G. Williamson, C. T. Foxon, and J. J. Harris, Phys. Rev. Lett. \textbf{62}, 1181 (1989).
\bibitem{ref39} K. G. Budden {\em The Waveguide Mode Theory of Wave Propagation\/} (Prentice-Hall, London, 1961).
\bibitem{ref40} I. Tolstoy, Proc.\ Symp.\ on Quasi-Optics, Microwave Research Institute Symposia Series Vol.\ XIV (Polytechnic Press, New York, 1960).
\bibitem{ref41} L. M. Brekhovskikh, {\em Waves in Layered Media\/} (Academic Press, New York, 1960).
\bibitem{ref42} Lord Rayleigh, {\em Theory of Sound}, 2nd ed.\ (Macmillan, London, 1896): par.\ 287; Phil.\ Mag.\ {\bf 20}, 1001 (1910); Phil.\ Mag.\ {\bf 27}, 100 (1914).
\bibitem{ref43} G. Timp, A. M. Chang, P. Mankiewich, R. Behringer, J. E. Cunningham, T. Y. Chang, and
R. E. Howard, Phys. Rev. Lett. \textbf{59}, 732 (1987); G. Timp, in Ref.\ \cite{ref13}; G. Timp, P. M. Mankiewich, P. DeVegvar, R. Behringer, J. E. Cunningham, R. E. Howard,
 H. U. Baranger, and J. K. Jain, Phys. Rev. B \textbf{39}, 6227 (1989).
\bibitem{ref44} A. M. Chang, K. Owusu-Sekyere, and T. Y. Chang, Solid State Comm. \textbf{67}, 1027 (1988).
\bibitem{ref45} C. J. B. Ford, T. J. Thornton, R. Newbury, M. Pepper, H. Ahmed, D. C. Peacock, D. A.
Ritchie, J. E. F. Frost, and G. A. C. Jones, Phys. Rev. B \textbf{38}, 8518 (1988); T. J. Thornton, {\em et al.\/} in Ref.\ \cite{ref12}.
\bibitem{ref46} M. L. Roukes, A. Scherer, S. J. Allen, Jr., H. G. Craighead, R. M. Ruthen, E. D. Beebe, and J. P. Harbison, Phys. Rev. Lett. \textbf{59}, 3011 (1987).
\bibitem{ref47} K. Ishibashi, Y. Takagaki, K. Garno, S. Namba, S. Ishida, K. Murase, Y. Aoyagi, and M. Kawabe, Solid State Comm. \textbf{64}, 573 (1987).
\bibitem{ref48} M. B\"{u}ttiker, Phys. Rev. B \textbf{38}, 9375 (1988).
\bibitem{ref49} H. van Houten, C. W. J. Beenakker, P. H. M. van Loosdrecht, T. J. Thornton, H. Ahmed,
M. Pepper, C. T. Foxon, and J. J. Harris, Phys. Rev. B \textbf{37}, 8534 (1988); R. J. Haug, A. H. MacDonald, P. Streda, and K. von Klitzing, Phys. Rev. Lett. \textbf{61}, 2797 (1988); S. Washburn, A. B. Fowler, H. Schmid, and D. Kern, Phys. Rev. Lett. \textbf{61}, 2801 (1988).
\bibitem{ref50} P. A. Lee, A. D. Stone, and H. Fukuyama, Phys. Rev. B \textbf{35}, 1039 (1987).
\bibitem{ref51} C. W. J. Beenakker and H. van Houten, Phys. Rev. B \textbf{37}, 6544 (1988).
\bibitem{ref52} Z. Tesanovic, M. V. Jaric, and S. Maekawa, Phys. Rev. Lett. \textbf{57}, 2760 (1986); Y. Isawa, Surf Sci. \textbf{170}, 38 (1986).
\bibitem{ref53} X. C. Xie and S. Das Sarma, Solid State Comm.\ {\bf 68}, 697 (1988).
\bibitem{ref54} J. H. F. Scott-Thomas, S. B. Field, M. A. Kastner, H. I. Smith, and D. A. Antoniadis,
Phys. Rev. Lett. \textbf{62}, 583 (1989); A. M. Chang, G. Timp, R. E. Howard, R. E. Behringer, P. M. Mankiewich, J. E. Cunningham, T. Y. Chang, and B. Chelluri, Superlattices and Microstructures \textbf{4}, 515 (1988).
\end{thebibliography}
\end{document}